\documentclass{iau}
\usepackage{graphicx}
\usepackage{subcaption}
\newcommand{\Teff}{\ensuremath{T_{\mathrm{eff}}}}              

\title[The age of Taurus - disc lifetimes] 
{The age of Taurus - environmental effects on disc lifetimes}

\author[J. M. Rees, T. Wilson, C. P. M. Bell, R. D. Jeffries \& T. Naylor]   
{J. M. Rees$^1$,
  T. Wilson$^1$, C. P. M. Bell$^2$, R. D. Jeffries$^3$ \and T. Naylor$^1$}

\affiliation{$^1$School of Physics, University of Exeter, Exeter, EX4 4QL, UK\\email: {\tt jon@astro.ex.ac.uk}\\
$^2$Dept. of Physics \& Astronomy, University of Rochester, Rochester, NY 14627-0171, USA \\
$^3$Astrophysics Group, Keele University, Staffordshire, ST5 5BG, UK}

\pubyear{2015}
\volume{314}  
\pagerange{?--?}
\jname{Young Stars \& Planets Near the Sun}
\editors{J. H. Kastner, B. Stelzer, \& S. A. Metchev, eds.}
\begin{document}

\maketitle

\begin{abstract}
Using semi-empirical isochrones, we find the age of the Taurus star-forming region to be 3-4 Myr. Comparing the disc fraction in Taurus to young massive clusters suggests discs survive longer in this low density environment. We also present a method of photometrically de-reddening young stars using $iZJH$ data. 
\keywords{stars: evolution, stars: pre-main sequence, stars: fundamental parameters}
\end{abstract}

\firstsection 
\section{Introduction}
Taurus is a low-density star-forming region containing primarily low-mass stars and so represents an ideal laboratory for studying the environmental effects on circumstellar disc lifetimes \cite[(Kenyon et al. 2008)]{Kenyon2008}. To investigate the impact of the low-density environment on the discs in Taurus, we used the Wide-Field Camera (WFC) on the 2.5m Isaac Newton Telescope (INT) on La Palma to obtain $griZ$ photometry of 40 fields in Taurus. Our fields are focused on the densest regions not covered by the Sloan Digital Sky Survey. The resultant INT-WFC survey mainly covers the L1495, L1521 and L1529 clouds.

We have augmented the WFC data with near-infrared $JHK$ data from 2MASS \cite[(Cutri et al. 2003)]{Cutri2003}.
To determine the age of Taurus we compare the semi-empirical isochrones discussed in \cite[Bell et al. (2013, 2014)]{Bell2013, Bell2014} to the observed colour-magnitude diagrams (CMDs). For a brief description of these isochrones see Bell et al. (these proceedings).  

\section{De-reddening}
\begin{figure}[b]
\begin{center}
\begin{subfigure}[Figure A]{0.45\textwidth}
\includegraphics[width=\textwidth, height=5.5cm]{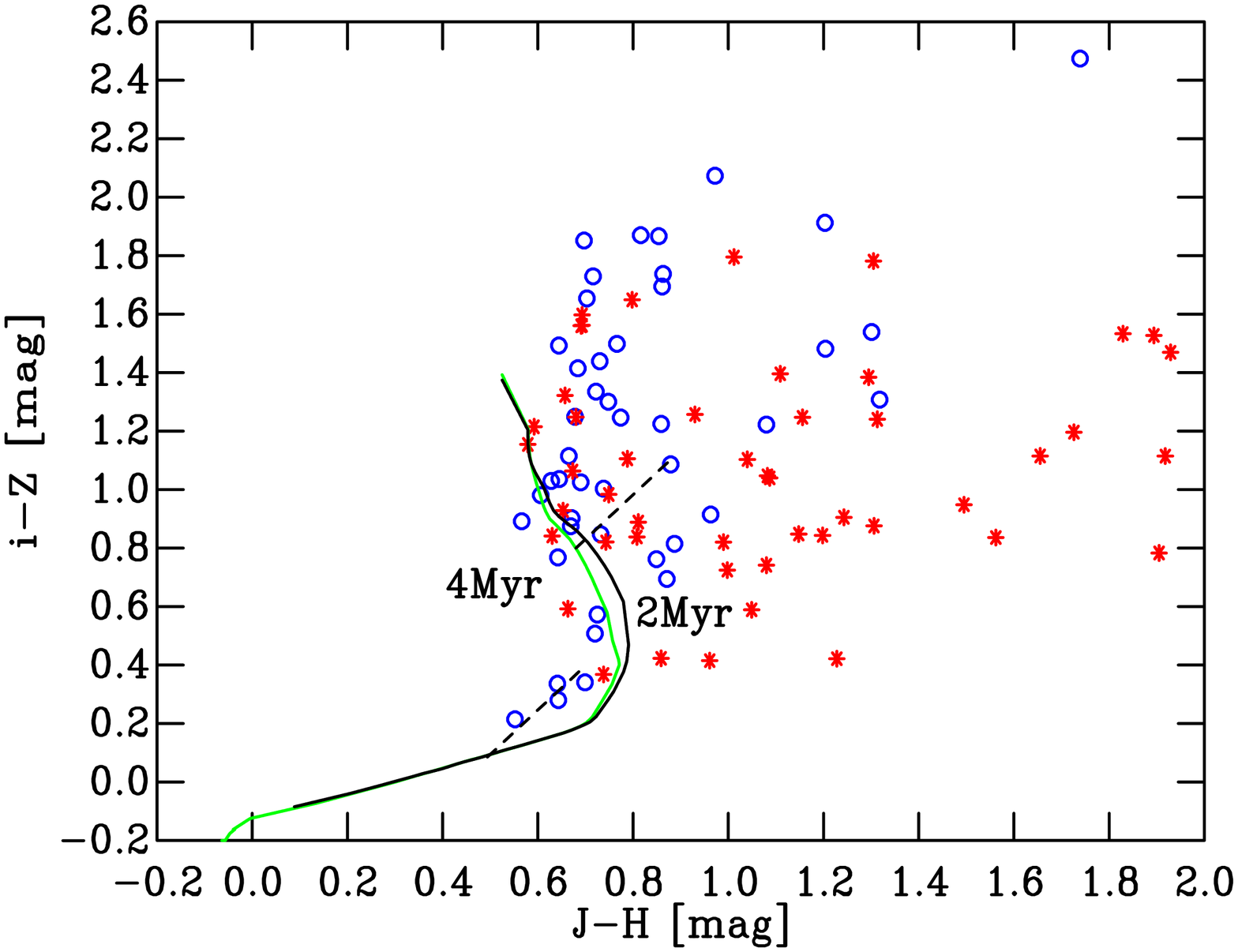}
\label{fig:izjh_ccd}
\end{subfigure}
\begin{subfigure}[Figure B]{0.45\textwidth}
\includegraphics[width=\textwidth, height=5.5cm]{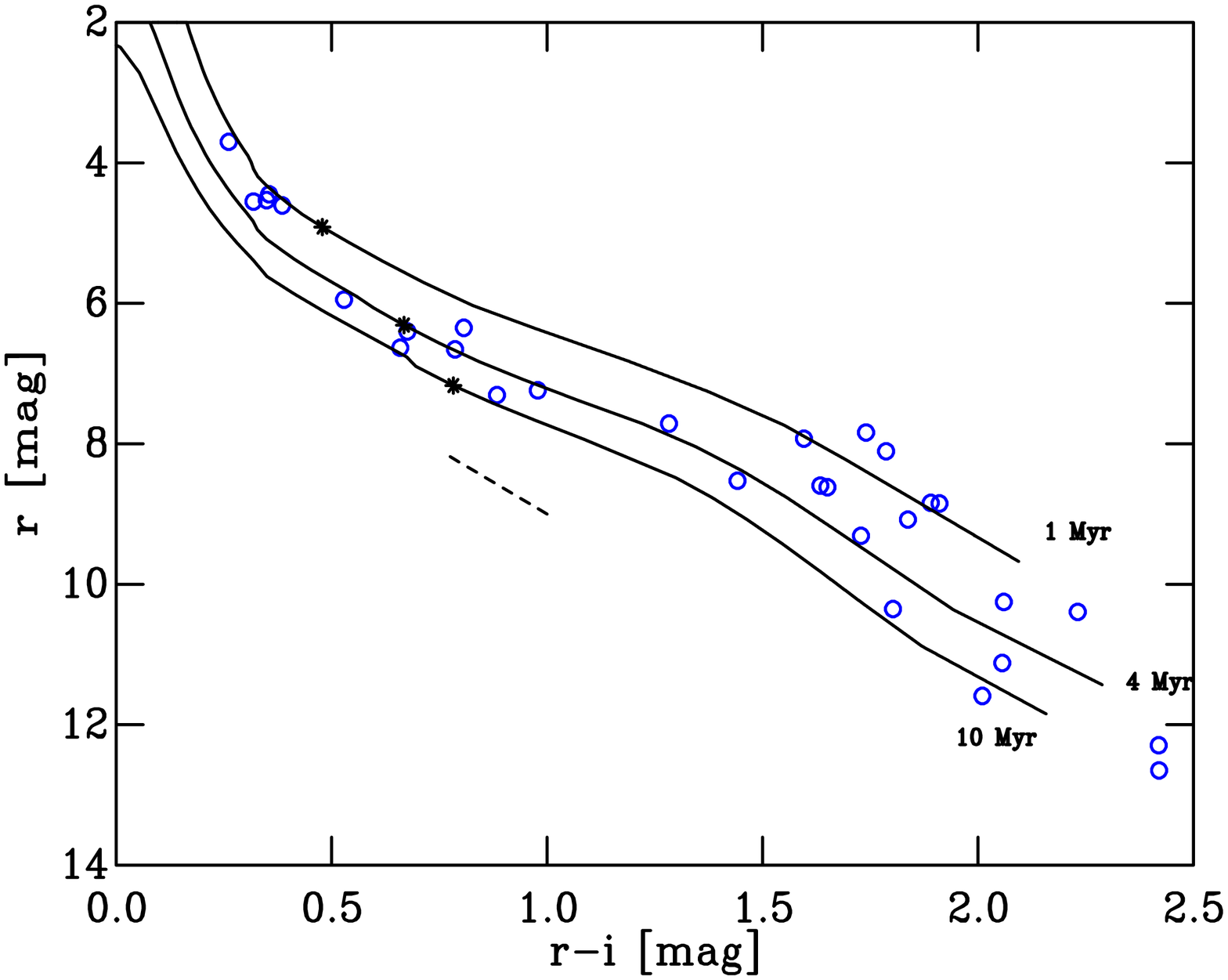}
\label{fig:age}
\end{subfigure}
\caption{\textbf{Left:} $i$-$Z$, $J$-$H$ diagram for Taurus members. Asterisks are Class II sources, open circles are Class III sources. Overlaid as solid lines are a 2 and 4\,Myr isochrone. The dashed lines are reddening vectors in this colour space. \textbf{Right:} $r$, $r$-$i$ diagram for Taurus members identified as Class III. Isochrones of 1, 4 and 10\,Myr are overlaid. Asterisks indicate the position of a theoretical star with mass 0.75M$_\odot$. The black dashed line shows a reddening vector for A$_V$ = 1 mag. }
\label{fig:izjh_age}
\end{center}
\end{figure}

The extinction in Taurus is spatially variable across the different clouds, and so we require a method of de-reddening the stars individually.
We have found that in an $i$-$Z$, $J$-$H$ colour-colour diagram the reddening vectors are almost perpendicular to the theoretical stellar sequence (Fig.\,\ref{fig:izjh_age}), whose position is almost independent of age, and so we can de-redden stars using photometry alone.
We construct a grid of models over a range of ages (1 to 10 Myr) and binary mass ratios (single star to equal mass binary).
We adopt the reddening law from \cite[Fitzpatrick (1999)]{Fitzpatrick1999}, apply it to the atmospheric models and fold the result through sets of filter responses to derive reddening coefficients in each photometric system.
It is well known that for a fixed value of E(B-V), extinction in a given filter will vary with \Teff\ (see e.g. \cite[Bell et al. 2013]{Bell2013}). To account for this we use extinction tables to redden the isochrones for a grid of E($B$-$V$) and \Teff\ values, and compare the reddened model grid to the data. We adopt a Bayesian approach and marginalise over binary mass, age and \Teff. We take the extinction values from the model with the highest likelihood, and use this to de-redden the star.

\section{Taurus}
Plotting the de-reddened Taurus members in the $r$, $r$-$i$ CMD, we notice that a significant fraction of the Class II objects appear much fainter than the primary locus. This is likely an accretion effect, and if we were to fit for the age of these members we would derive an age that is erroneously old. To avoid this effect, we fit only the Class III sources. We note that those Class II sources that are not scattered below the sequence lie coincident with the Class III sources, and thus we believe the age derived from the Class III sources alone will be representative of the overall age.
We plot our de-reddened Taurus members in an $r$, $r$-$i$ CMD to fit for the age (Fig.\,\ref{fig:izjh_age}). We find that isochrones of 3-4 Myr (older than is commonly quoted in the literature) trace the observed stellar sequence well. To ensure consistency with the \cite[Bell et al. (2013)]{Bell2013} age scale we compare the position of a theoretical star with a mass of 0.75 M$_\odot$ to the middle of the observed sequence. We find consistency with the overall isochrone fitting, with an age of 3-4 Myr still providing a good fit. 
With a robust age for Taurus we then examined the disc fraction. Taurus has a disc fraction of 69\% \cite[(Luhman et al. 2010)]{Luhman2010}. If we compare this to the other clusters in \cite[Bell et al. (2013)]{Bell2013}, which are on the same age scale, we find that Taurus has the largest disc fraction in the sample, significantly higher than the group of young (2 Myr), massive clusters, suggesting that discs may have survived longer in the low density environment present in Taurus.

\end{document}